\documentclass[12pt]{article}
\usepackage{amsfonts,amssymb}

\def\C{\mathbb{C}}

\def\bq{ \begin{equation} }
\def\eq{ \end{equation} }
\def\ben{ \begin{eqnarray} }
\def\en{ \end{eqnarray} }

\def\frac#1#2{{#1\over #2}}

\def\on#1#2{\mathop{\vbox{\ialign{##\crcr\noalign{\kern2pt}
$\scriptstyle{#2}$\crcr\noalign{\kern2pt\nointerlineskip}
\kern-2pt$\hfil\displaystyle{#1}\hfil$\crcr}}}\limits}

\begin{document}

\title{A family of integrable evolution equations of third order}
\author{M. Babela and  A. Odesskii}
   \date{}
\vspace{-20mm}
   \maketitle
\vspace{-7mm}
\begin{center}
Department of Mathematics and Statistics \\
Brock University \\
1812 Sir Isaac Brock Way, St. Catharines, ON, L2S 3A1 Canada \\[1ex]
e-mail: \\
\texttt{aodesski@brocku.ca}
\end{center}

\medskip

\begin{abstract}

We construct a family of integrable equations of the form $v_t=f(v,v_x,v_{xx},v_{xxx})$ such that $f$ is a transcendental function  in $v,v_x,v_{xx}$. This 
family is related to the Krichever-Novikov equation by a differential substitution. Our construction of integrable equations and the corresponding differential substitutions 
involves geometry of a family of genus two curves and their Jacobians.

\medskip


\medskip

\end{abstract}

\newpage

\tableofcontents

\newpage

\section{Introduction}

We address the problem of classification of integrable equations of the form 
\begin{equation}   \label{int}
 v_t=f(v,v_x,v_{xx},v_{xxx})
\end{equation}
where $v=v(x,t)$ and indexes stand for partial derivatives. Using the 
so-called symmetry approach (see \cite{fok,MSS} and references therein), it was proven \cite{IS} that any integrable equation (\ref{int})  must depend on $v_{xxx}$ in one of the following ways:

\begin{equation}   \label{int1}
v_t=f_1v_{xxx}+f_2
\end{equation}

\begin{equation}   \label{int2}
v_t=\frac{1}{(f_1v_{xxx}+f_2)^2}+f_3
\end{equation}

\begin{equation}   \label{int3}
v_t=\frac{2f_1v_{xxx}+f_2}{\sqrt{f_1v_{xxx}^2+f_2v_{xxx}+f_3}}+f_4
\end{equation}

where $f_i$ are functions in $v,v_x,v_{xx}$ and $f_2^2-4f_1f_3\ne 0$ in (\ref{int3}). In particular, $f$ must be rational  or  algebraic function in $v_{xxx}$. Integrable equations of the form (\ref{int1}) 
were classified in \cite{MSS}. It was shown  that any such equation has a form

\begin{equation}   \label{int11}
 v_t=\frac{v_{xxx}+g_1v_{xx}^3+g_2v_{xx}^2+g_3v_{xx}+g_4}{(g_5v_{xx}^2+g_6v_{xx}+g_7)^{3/2}}+g_8
\end{equation}

where $g_1,...,g_8$ are functions in $v,v_x$. Thus, in any equation of this class, $f$ is an algebraic function in both variables $v_{xx},~v_{xxx}$. Moreover, this is the case for all 
known examples of integrable equations (\ref{int}). Equations of the form (\ref{int2}), (\ref{int3}) have not been classified yet but in the paper \cite{her} all equations of the 
form (\ref{int3}) with $f_1=0$ were listed, and in all these equations $f$ is an algebraic function in $v_{xx}$. 

One may believe that for any integrable equation (\ref{int}) the r.h.s.\  must be an algebraic function in both variables $v_{xx},~v_{xxx}$. We will show, however, that this is not correct. In 
this paper we construct a family of equations from the class (\ref{int2}) such that $f$ is not algebraic in $v_{xx}$. These equations have a form (\ref{int}) with

$$f(v_0,v_1,v_2,v_3)=-\frac{1}{2(1+G_{v_0}v_1+G_{v_1}v_2+G_{v_2}v_3)^2G_{v_2}}+R(v_0)v_1$$

where $G=G(v_0,v_1,v_2)$ is a transcendental function  in $v_0,v_1,v_2$.

Recall that equations of the form (\ref{int}) are said to be equivalent if they are connected by an invertible point or contact transformation of the form 

\begin{equation}   \label{cont}
\bar{t}=t,~~~~~ \bar{x}=x+\lambda(v,v_x),~~~\bar{v}=\mu(v,v_x)
\end{equation}

where functions $\lambda(v_0,v_1)$ and $\mu(v_0,v_1)$ satisfy the relation

\begin{equation}   \label{cont1}
 (1+v_1\lambda_v)\mu_{v_1}=v_1\mu_v\lambda_{v_1}
\end{equation}

Any such transformation does not change the forms of (\ref{int1}) - (\ref{int11}). In particular, transformation  (\ref{cont}) leads to fractional linear transformation 
of the variable $v_{xx}$ of the form $v_{xx}\to \frac{Av_{xx}+B}{Cv_{xx}+D}$ where $AD-BC\ne 0$ and $A,B,C,D$ are certain functions in $v,v_x$. Therefore, if the r.h.s.\ of (\ref{int}) 
is algebraic in $v_{xx}$, it will stay algebraic after arbitrary transformation (\ref{cont}).

There exists, however, a more sophisticated and highly nontrivial way of relations between integrable equations: the so-called differential substitutions (see \cite{sok1,sok2,sok3,MSS}). In 
this paper we obtain our family of integrable equations via differential substitutions of the form 
$$\bar{t}=a^3t,~~~y=ax+aG(v,v_x,v_{xx}),~~~
u=Q(v,v_x,v_{xx})$$
into the so-called Krichever-Novikov equation \cite{KN}
$$u_{\bar{t}}=u_{yyy}-\frac{3}{2}\frac{u_{yy}^2}{u_y}+\frac{c_1+c_2u+c_3u^2+c_4u^3+c_5u^4}{u_y}.$$

Notice, that the functions $G,Q$ are both transcendental and must satisfy a complicated system of nonlinear PDEs  (\ref{cons1})-(\ref{cons4}). We believe that we have found a general 
solution to this system up to transformations (\ref{cont}), and assuming that $G$ is not a fractional linear function in $v_{xx}$ (probably some other degenerate cases should be excluded), 
but we don't prove this here. 

In the next section we describe our construction of a family of integrable equations and differential substitution related it to the Krichever-Novikov equation. Proofs will be outlined in 
the last section. 

Some aspects of our construction can be interpreted in terms of the geometry of elliptic and genus two algebraic curves (see Remarks 3, 4 in the next section). Roughly speaking, $v_0$ can be 
regarded as a function on the moduli space of genus two curves and $v_1,v_2,G$ can be regarded as functions on Jacobians of these curves.

Much work is still needed to classify integrable equations of the form (\ref{int2}), (\ref{int3}). We hope that this paper sheds some light on the nature of their r.h.s.\ 
as functions of $v,v_x,v_{xx}$.

\section{A family of integrable equations}

Let us fix constants $c_1,...,c_5,h,b\in\C$. Define a function $Y(t,v_0)$ by

\begin{equation}   \label{Y}
 Y(t,v_0)=\sqrt{(v_0-h)(t-v_0)(t-h)(c_1+c_2t+c_3t^2+c_4t^3+c_5t^4)}
\end{equation}

We will often omit the second argument and write $Y(t)$ for $Y(t,v_0)$.

Define also functions $P_1=P_1(v_0,v_1,v_2),~P_2=P_2(v_0,v_1,v_2)$ implicitly by 

\begin{equation} \begin{array}{cc}  \label{P}
\int\limits_{b}^{P_1}\frac{(t-h)d t}{Y(t)}+\int\limits_{b}^{P_2}\frac{(t-h)dt}{Y(t)}=-\frac{2(v_0-h)}{v_1}, \\
\int\limits_{b}^{P_1}\frac{(t-h)d t}{(t-v_0)Y(t)}+\int\limits_{b}^{P_2}\frac{(t-h)dt}{(t-v_0)Y(t)}=\frac{4(v_0-h)v_2}{v_1^3}-\frac{6}{v_1}
\end{array}\end{equation}

{\bf Proposition 1.} The equation $v_t=F(v,v_x,v_{xx},v_{xxx})$  is integrable if 

\begin{equation}   \label{F}
F(v_0,v_1,v_2,v_3)=-\frac{1}{2(1+G_{v_0}v_1+G_{v_1}v_2+G_{v_2}v_3)^2G_{v_2}}+R(v_0)v_1
\end{equation}

where 

\begin{equation} \begin{array}{cc}  \label{GR}
G(v_0,v_1,v_2)=\frac{2(v_0-h)}{v_1}+\int\limits_{b}^{P_1}\frac{(v_0-h)d t}{Y(t)}+\int\limits_{b}^{P_2}\frac{(v_0-h)dt}{Y(t)}, \\ [7mm]
R(v_0)=\frac{1}{8}(c_2+2c_3h+3c_4h^2+4c_5h^3)+\frac{c_1+c_2h+c_3h^2+c_4h^3+c_5h^4}{2(v_0-h)}
\end{array}\end{equation}

{\bf Remark 1.} The constant $b$ in (\ref{P}), (\ref{GR}) is not essential. Changing of $b$ will lead to addition of certain functions in $v_0$ to the l.h.s.\ of (\ref{P}) 
and to the r.h.s.\ of (\ref{GR}) which can be absorbed by a point transformation of the form

\begin{equation}  \begin{array}{cc} \label{point}
x\to x+q(v), \\
v\to v
\end{array}\end{equation}

{\bf Remark 2.} The point transformation $v\to \frac{\alpha v+\beta}{\gamma v+\delta}$, where $\alpha,\beta,\gamma,\delta$ are constants, does not change the form of our equation. These 
transformations can be used to simplify our equation. For example, if the coefficients of the polynomial $c_1+c_2t+c_3t^2+c_4t^3+c_5t^4$ are generic, we can put this polynomial to 
canonical form $4t^3-g_2t-g_3$.

{\bf Proposition 2.} The equation 

\begin{equation}   \label{eq}
v_t=F(v,v_x,v_{xx},v_{xxx})
\end{equation}

 defined by (\ref{F}), (\ref{GR}) is connected with the Krichever-Novikov equation 

\begin{equation}   \label{KN}
u_{\bar{t}}=u_{yyy}-\frac{3}{2}\frac{u_{yy}^2}{u_y}+\frac{c_1+c_2u+c_3u^2+c_4u^3+c_5u^4}{u_y} 
\end{equation}

by differential substitution

\begin{equation} \begin{array}{ccc}  \label{subs}
\bar{t}=a^3t,\\
y=ax+aG(v,v_x,v_{xx}),\\
u=Q(v,v_x,v_{xx})
\end{array}\end{equation}

where $a=\frac{\sqrt[4]{3}}{2\sqrt[4]{2}}\sqrt[4]{c_1+c_2h+c_3h^2+c_4h^3+c_5h^4}$, $G$ is defined by (\ref{GR}) and the function $Q$ is defined by 

\begin{equation}   \label{Q}
2\int\limits_{b}^{Q}\frac{d t}{Z(t)}=
\int\limits_{b}^{P_1}\frac{d t}{Z(t)}+\int\limits_{b}^{P_2}\frac{d t}{Z(t)} 
\end{equation}

with $Z(t)=\sqrt{c_1+c_2t+c_3t^2+c_4t^3+c_5t^4}$. Recall that functions $P_1,P_2$ are defined by (\ref{P}). 

{\bf Remark 3.} Consider a family of genus two curves 
$$C=\{(x,y)\in\C^2;~y^2=(v_0-h)(x-v_0)(x-h)(c_1+c_2x+c_3x^2+c_4x^3+c_5x^4)\}.$$
Recall that $\frac{dx}{y},~\frac{xdx}{y}$ are two holomorphic differentials on $C$ and $\frac{(x-h)dx}{(x-v_0)y}$ is a meromorphic differential on $C$. Therefore, $P_1,P_2$ can be 
regarded as points of the curve $C$. Moreover, equations (\ref{P}), (\ref{GR}) for $P_1,P_2$ and $G$ can be regarded as a Jacobi inversion problem on the Jacobian of the curve $C$.

{\bf Remark 4.} Consider a family of elliptic curves 
$$E=\{(x,z)\in\C^2;~z^2=c_1+c_2x+c_3x^2+c_4x^3+c_5x^4\}.$$
Equation (\ref{Q}) can be written in terms of the addition law on $E$ as $2[Q]=[P_1]+[P_2]$, where $[Q]$ (resp. $[P_1],~[P_2]$) is a point on $E$ with $x=Q$ (resp. $x=P_1,~x=P_2$). It is known that 
(\ref{Q}) is equivalent to an algebraic relation between $Q,P_1,P_2$. Explicitly, we have
\begin{equation} \begin{array}{cc}  \label{algQ}
 \frac{c_2^2+4(c_2c_3-2c_1c_4)Q+2(2c_3^2-8c_1c_5-c_2c_4)Q^2+4(c_3c_4-2c_2c_5)Q^3+c_4^2Q^4}{c_1+c_2Q+c_3Q^2+c_4Q^3+c_5Q^4}= \\[4mm]
=4\frac{2c_1+c_2P_1+c_2P_2+c_3P_1^2+c_3P_2^2+c_4P_1^2P_2+c_4P_1P_2^2+2c_5P_1^2P_2^2-2Z(P_1)Z(P_2)}{(P_1-P_2)^2}
\end{array}\end{equation}

{\bf Remark 5.} Equations (\ref{P}), (\ref{GR}) can be written in terms of multi-dimensional $\zeta$ and $\wp$-functions associated with the curve $C$ (see \cite{BEL} and references therein). 
We have extensively used results and formulas from \cite{BEL} in our intermediate computations.

\section{Proofs}

It is known that if an equation (\ref{int}) is connected with an integrable equation by a differential substitution, then the equation (\ref{int}) is also integrable \cite{sok2}. 
It is also known that the Krichever-Novikov equation is integrable \cite{KN}. Therefore, Proposition 1 follows from Proposition 2. Let us prove Proposition 2.

After substitution of (\ref{subs}) into the Krichever-Novikov equation (\ref{KN}) and by virtue of (\ref{eq}) we obtain a large expression\footnote{We use notations $v_0=v,v_i=v_{x^i}$.} 
in $v_0,...,v_5$ which is equal to zero 
if and only if the equation (\ref{eq}), with $F$ given by (\ref{F}), is related with the Krichever-Novikov equation (\ref{KN}) by (\ref{subs}). Moreover, this expression is a polynomial in 
$v_3,v_4,v_5$. Splitting by $v_3,v_4,v_4$ one can check that this polynomial is equal to zero if and only if the following equations hold:

\begin{equation}   \label{cons1}
v_1(Q_{v_0}G_{v_2}-Q_{v_2}G_{v_0})+v_2(Q_{v_1}G_{v_2}-Q_{v_2}G_{v_1}) -Q_{v_2}=0
\end{equation}

\begin{equation}   \label{cons2}
2Q_{v_2}G_{v_2}(Q_{v_1v_2}G_{v_2}-G_{v_1v_2}Q_{v_2})+G_{v_2v_2}Q_{v_2}(Q_{v_2}G_{v_1}+Q_{v_1}G_{v_2})-2Q_{v_2v_2}Q_{v_1}G_{v_2}^2=0
\end{equation}

\begin{equation} \begin{array}{ccc}  \label{cons3}
\frac{3}{16}G_{v_2}^6(c_1+c_2h+c_3h^2+c_4h^3+c_5h^4)(c_1+c_2Q+c_3Q^2+c_4Q^3+c_5Q^4)-\\[5mm]
-2Q_{v_2}^2G_{v_2}^4R(v_0)-2v_1^2Q_{v_2}G_{v_2}^4(Q_{v_1}G_{v_2}-Q_{v_2}G_{v_1})\frac{dR(v_0)}{dv_0}+\\[5mm]
+2Q_{v_2v_2v_2}Q_{v_2}G_{v_2}^2-2G_{v_2v_2v_2}Q_{v_2}^2G_{v_2}+3G_{v_2v_2}^2Q_{v_2}^2-3Q_{v_2v_2}^2G_{v_2}^2=0
\end{array}\end{equation}

\begin{equation}  \begin{array}{cc} \label{cons4}
\frac{3Q_{v_2}^2G_{v_2}^2}{Q_{v_1}G_{v_2}-Q_{v_2}G_{v_1}}=3v_2Q_{v_2}G_{v_2}^2+v_1Q_{v_2}G_{v_2v_2}+v_1G_{v_2}(3Q_{v_1}G_{v_2}-Q_{v_2}G_{v_1})+\\[5mm]
+v_1v_2Q_{v_2}(G_{v_2v_2}G_{v_1}-G_{v_1v_2}G_{v_2})+v_1^2Q_{v_2}(G_{v_2v_2}G_{v_0}-G_{v_0v_2}G_{v_2})
\end{array}\end{equation}

On the other hand, differentiating the relations (\ref{GR}), (\ref{P}) by $v_0,v_1,v_2$ one gets the following system of PDEs for $G,P_1,P_2$:
\begin{equation}  \begin{array}{cccccc} \label{PDE1}
G_{v_2}=\frac{4(v_0-h)^2(P_1-v_0)(P_2-v_0)}{v_1^3(P_1-h)(P_2-h)},\\[4mm]
G_{v_1}=
\frac{2(v_0-h)(2P_1P_2-(3v_0-h)(P_1+P_2)+4v_0^2-2hv_0)}{v_1^2(P_1-h)(P_2-h)}-\frac{12v_2(v_0-h)^2(P_1-v_0)(P_2-v_0)}{v_1^4(P_1-h)(P_2-h)} \\[4mm]
P_{1,v_2}=-\frac{4(v_0-h)(P_1-v_0)(P_2-v_0)Y(P_1)}{v_1^3(P_1-P_2)(P_1-h)},~P_{2,v_2}=\frac{4(v_0-h)(P_1-v_0)(P_2-v_0)Y(P_2)}{v_1^3(P_1-P_2)(P_2-h)},\\[4mm]
P_{1,v_1}=\frac{2(6v_2(v_0-h)(P_2-v_0)-v_1^2(3P_2-4v_0+h))(P_1-v_0)Y(P_1)}{v_1^4(P_1-P_2)(P_1-h)},\\[4mm]
P_{2,v_1}=-\frac{2(6v_2(v_0-h)(P_1-v_0)-v_1^2(3P_1-4v_0+h))(P_2-v_0)Y(P_2)}{v_1^4(P_1-P_2)(P_2-h)},\\[4mm]
P_{1,v_0}=-\frac{Y(P_1)}{v_1^3(v_0-h)(P_1-P_2)}(v_1^3(P_2-h)G_{v_0}+v_1^2(P_2-h)-2v_2(v_0-h)(P_2-v_0)),\\[4mm]
P_{2,v_0}=\frac{Y(P_2)}{v_1^3(v_0-h)(P_1-P_2)}(v_1^3(P_1-h)G_{v_0}+v_1^2(P_1-h)-2v_2(v_0-h)(P_1-v_0)).                   
\end{array}\end{equation}

Equation (\ref{Q}) shows that 

\begin{equation}   \label{S}
 Q=S(P_1,P_2)
\end{equation}

where

\begin{equation}   \label{PDE2}
S_{P_1}=\frac{Z(S)}{2Z(P_1)},~~~S_{P_2}=\frac{Z(S)}{2Z(P_2)}. 
\end{equation}

Recall that $Z(t)=\sqrt{c_1+c_2t+c_3t^2+c_4t^3+c_5t^4}$.

Now let us outline computations needed to prove relations (\ref{cons1}) - (\ref{cons4}). To prove (\ref{cons1}) we start by substituting (\ref{S}) for $Q$. After that we check 
that (\ref{cons1}) becomes an identity by virtue of (\ref{PDE1}). To prove (\ref{cons2}) we proceed with the same steps but in the end we also need to substitute (\ref{PDE2}). 
Relations (\ref{cons3}), (\ref{cons4}) can be proven in the same way but for (\ref{cons3}) we need algebraic relation (\ref{algQ}) as well.

\end{document}